\input amstex
\input amsppt.sty
\loadbold
\pagewidth{18cm}\pageheight{23cm}
\CenteredTagsOnSplits
\define\const{\operatorname{const}}
\topmatter
\title
Dynamical systems accepting the normal shift.
\endtitle
\author
Boldin A.Yu. and Sharipov R.A.
\endauthor
\address
Department of Mathematics,
Bashkir State University, Frunze str. 32, 450074 Ufa,
Russia
\endaddress
\email
root\@bgua.bashkiria.su
\endemail
\date
March 9, 1993.
\enddate
\abstract
Classical Bianchi-Lie, B\"acklund and Darboux transformations are
considered. Their generalizations for the dynamical systems are
discussed. For the transformation being the generalization of
the normal shift the special class of dynamical systems is
defined. The effective criterion for separating such systems in
form of partial differential equations is found.
\endabstract
\endtopmatter
\document
\head
1. Introduction
\endhead
     B\"acklund transformations which are well known to the
specialists in integrable nonlinear equations originally first
arose in the works of classics of differential geometry in last
century. Let's remind some of their results. Consider
2-dimensional surface $S$ in $\Bbb R^3$ with the standard scalar
product. Let us map each point $M$ of the surface $S$ onto the
point $M'$ of the other surface $S'$ so that the following
conditions are fulfilled
\roster
\item The distance $\left|MM'\right|$ is constant and it is equal
      to $R$;
\item The tangent planes $\tau$ and $ \tau'$ of the surfaces $S$
      and $S'$ in the points $M$ and $M'$ are orthogonal:
      $\angle\left(\tau,\tau'\right)=90^\circ$;
\item The segment $MM'$ lies in both tangents planes $\tau$ and
      $\tau'$ of the surfaces $S$ and $S'$ respectively.
\endroster
Bianchi \cite{1} showed that if $S$ is the surface of constant
negative Gaussian curvature $K=-1/R^2$ then in assumptions
\therosteritem{1}--\therosteritem{3} the second surface $S'$
also is the surface of constant negative Gaussian curvature
$K'=-1/R^2$. Lie \cite{2} strengthened Bianchi's result having
shown that construction \therosteritem{1}--\therosteritem{3}
can be realized only on the surface $S$ of constant negative
Gaussian curvature $K=-1/R^2$. \par
     B\"acklund \cite{3} generalized the construction
\therosteritem{1}--\therosteritem{3} having substituted
condition \therosteritem{2} by more weak condition of constancy
of the angle between the tangent planes $\tau$ and $\tau'$. Such
kind of transformation also is defined only on the surfaces of
constant negative curvature and it leads to the surfaces $S'$ of
the same curvature.
     Darboux \cite{4} offered further generalization of this
construction having substituted condition \therosteritem{3}
by condition of constancy of the angles between line $MM'$ and
both tangent planes $\tau$ and $\tau'$. However thereby some
differences there appeared: Darboux transformation is realized
on the surfaces where some linear combination of Gaussian and
mean curvatures is constant. The result of transformation is the
surface where some other linear combination of Gaussian and mean
curvatures is constant. \par
     Differential equations which are obtained in considering
the transformations \therosteritem{1}--\therosteritem{3} as
well as their generalizations became the objects of the numerous
investigations. For the modern state of such investigations one
can enquire the monographs \cite{5} and \cite{6}. The
generalization of the above transformations for the
multidimensional spaces and submanifolds in them is made in
papers  \cite{7}, \cite{8} and \cite{9}. The generalizations for
the submanifolds in the arbitrary Riemannian manifold of constant
sectional curvature is considered in \cite{10} and \cite{11}.
\par
     The normal shift (or the Bonnet transformation) is some
particular case in the general Darboux construction. In this case
tangent planes $\tau$ and $\tau'$ are parallel while the segment
$MM'$ is orthogonal to them both. This is degenerate particular
case since it can be realized on any surface imposing no
limitations for its curvatures. \par
     In this paper we discuss the generalization of the
Bianchi-Lie, B\"acklund and Darboux transformation for the
spacially anisotropic case substituting the straight line segment
or geodesic segment $MM'$ by the segment of trajectory for some
dynamical system. Let's consider the following dynamical system
in $\Bbb R^n$
$$
\ddot\bold r=\bold F(\bold r,\dot\bold r)\tag1.1
$$
This is the natural second order dynamical system defined by the
force $\bold F$. For each point $M$ on some submanifold
$S\subset\Bbb R^n$ let us consider the particle starting from
it with some initial velocity $\bold v$.
In the end of some time interval (same for all particles) these
particles form some other submanifold $S'\subset\Bbb R^n$. In
most general form the problem may be stated as follows: what kind
of limitations for the dynamical system \thetag{1.1} itself and
for the choice of submanifold $S$ and initial velocities of
particles on it arise if one require the angles defining the
mutual arrangement of tangent spaces $\tau$, $\tau'$ and particle
trajectories to be constant. Such problem has a lot of possible
specializations one of which leading to the meaningful results is
considered below. \par
\head
2. The normal shift along the dynamical system.
\endhead
     Let's consider the dynamical system \thetag{1.1} in
Euclidean space $\Bbb R^n$ with the standard scalar product. Let
$S$ be the submanifold of codimension 1 in $\Bbb R^n$ and let
$\bold n$ be the vector field of unit normal vectors on $S$. We
direct the initial velocity of particles along $\bold n$
defining the modulus of velocity as some smooth function $v=v(M)$
on the submanifold $S$. Then the dynamical system \thetag{1.1}
produces the family of submanifolds $S_t$
$$
f_t:S\longrightarrow S_t\tag2.1
$$
together with the diffeomorphisms $f_t$ binding $S$ with $S_t$.
\par
\definition{Definition 1} Each transformation $f=f_t$ of the
family \thetag{2.1} is called the normal shift along the
dynamical system \thetag{1.1} if each  trajectory of \thetag{1.1}
crosses each submanifold $S_t$ along its normal vector $\bold n$.
\enddefinition
\definition{Definition 2} Dynamical system \thetag{1.1} is called
the dynamical system accepting the normal shift of submanifolds
of codimension 1 if for any submanifold S of codimension 1 there
is the function $v=v(M)$ on $S$ such that the transformation
\thetag{2.1} defined by the system \thetag{1.1} and the initial
velocity function $\left|\bold v(M)\right|=v(M)$ is the
transformation of normal shift.
\enddefinition
     Note  that the transformation \thetag{2.1} may implement the
normal shift for some particular submanifolds even when they do
not satisfy the definition 2. Dynamical systems accepting the
normal shift for arbitrary submanifold form the special class of
dynamical systems narrow enough to be described in much details.
In the following two sections we consider such dynamical systems
in  $\Bbb R^2$ and derive the partial differential equation for
the force function F of them. \par
\head
3. Dynamical systems in $\Bbb R^2$ accepting the normal shift.
\endhead
     Let's consider the second order dynamical system
\thetag{1.1} in Euclidean space $\Bbb R^2$ with the standard
scalar product. Phase space of the system \thetag{1.1} is
four-dimensional in this case. The locus of points where
$\bold v=\dot\bold r=0$ is the two-dimensional plane in it.
Everywhere out of this locus we define the unit vector
$$
\bold N=\bold N(\bold v)=\frac{\bold v}{|\bold v|}\tag3.1
$$
and the unit vector $\bold M(\bold v)$ perpendicular to
$\bold N(\bold v)$. Using \thetag{3.1} the right hand side of
\thetag{1.1} can be rewritten as follows
$$
\bold F(\bold r,\bold v)=A(\bold r,\bold v)\bold N(\bold v)
+B(\bold r,\bold v)\bold M(\bold v)\tag3.2
$$
Cartesian components of $\bold N(\bold v)$ and $\bold M(\bold v)$
satisfy the following differential equations
$$
\frac{\partial N^k}{\partial v^i}=\frac{M_iM^k}{|\bold v|}
\hskip 10em
\frac{\partial M^k}{\partial v^i}=-\frac{M_iN^k}{|\bold v|}
\tag3.3
$$
Gradients of the functions $A(\bold r,\bold v)$ and $B(\bold r,
\bold v)$ in \thetag{3.2} also can be expressed in terms of
components of $\bold N(\bold v)$ and $\bold M(\bold v)$
$$
\xalignat 3
\frac{\partial A}{\partial r^i}&=\alpha_1N_i+\alpha_2M_i &
\frac{\partial A}{\partial v^i}&=\alpha_3N_i+\alpha_4M_i\tag3.4
\\
\frac{\partial B}{\partial r^i}&=\beta_1N_i+\beta_2M_i &
\frac{\partial B}{\partial v^i}&=\beta_3N_i+\beta_4M_i\tag3.5
\endxalignat
$$
For the dynamical system \thetag{1.1} with the right hand side
\thetag{3.2} we consider the Cauchy problem with the initial data
depending on some scalar parameter $s$
$$
\left.\bold r(t,s)\right|_{t=0}=\bold r(s)\hskip 10em
\left.\partial \bold r(t,s)\right|_{t=0}=\bold v(s)\tag3.6
$$
Let $\boldsymbol\tau(t, s)$ be the derivative of $\bold r(t, s)$
by the parameter $s$. Differentiating \thetag{1.1} we derive the
following equation for the components of the vector
$\boldsymbol\tau(t, s)$
$$
\ddot\tau^k=\frac{\partial F^k}{\partial r^i}\tau^i+
\frac{\partial F^k}{\partial v^i}\dot\tau^i\tag3.7
$$
Using the expression \thetag{3.2}, the differential equations
\thetag{3.3} and the formulae \thetag{3.4} and \thetag{3.5} we my
write the partial derivatives in \thetag{3.7} in the following
form
$$
\align
\frac{\partial F^k}{\partial r^i}&=\left(\alpha_1N_i+\alpha_2M_i
\right) N^k+\left(\beta_1N_i+\beta_2M_i\right)M^k\tag3.8 \\
\split
\frac{\partial F^k}{\partial v^i}&=\alpha_3N_iN^k+\left(\alpha_4-
\frac  B{|\bold v|}\right)M_iN^k+\\
&\hskip 10em+\beta_3N_iM^k+\left(\beta_4+\frac A{|\bold v|}
\right)M_iM^k
\endsplit
\tag3.9
\endalign
$$
Time derivative for the vector $\bold v$ is defined by the
equation \thetag{1.1}. For the length of this vector
then we have
$$
\partial_t|\bold v|=A\hskip 10em\partial_t\left(|\bold v|^{-1}
\right)=-\frac{A}{|\bold v|^2}
\tag3.10
$$
For the derivatives of $A$ and $B$ in \thetag{3.4} and
\thetag{3.5} we obtain
$$
\gathered
\partial_t A=\alpha_1|\bold v|+\alpha_3A+\alpha_4B \\
\partial_tB=\beta_1|\bold v|+\beta_3A+\beta_4B
\endgathered
\tag3.11
$$
Taking into account \thetag{3.4} and \thetag{3.5} from
\thetag{3.3} and \thetag{3.10} we derive the formulae for the
time derivatives of $\bold N$ and $\bold M$
$$
\partial_t\bold N=\frac{B\bold N}{|\bold v|}\hskip 10em
\partial_t\bold M=-\frac{B\bold N}{|\bold v|}\tag3.12
$$
\par
     Time dynamics of the vector $\boldsymbol\tau$ due to
\thetag{3.7} and the relationships \thetag{3.12} determine the
dynamics of the scalar products $\left<\boldsymbol\tau,
\bold N\right>$ and $\left<\boldsymbol\tau, \bold M\right>$.
Let's denote them as follows
$$
\left<\boldsymbol\tau,\bold N\right>=\varphi\hskip 10em
\left<\boldsymbol\tau, \bold M\right>=\phi\tag3.13
$$
Differentiating \thetag{3.13} and taking into account
\thetag{3.12} we obtain
$$
\gathered
\left<\partial_t\boldsymbol\tau,\bold N\right>=\partial_t\varphi-
\frac B{|\bold v|}\psi \\
\left<\partial_t\boldsymbol\tau,\bold M\right>=\partial_\psi+
\frac B{|\bold v|}\varphi
\endgathered
\tag3.14
$$
Differentiating the left hand sides in \thetag{3.14} we get
$$
\align
\split
\partial_t\left<\partial_t\boldsymbol\tau,\bold N\right>&=
\left<\partial_{tt}\boldsymbol\tau,\bold N\right>+
\left<\partial_t\boldsymbol\tau,\partial_t\bold N\right>=\\
&=\left<\partial_{tt}\boldsymbol\tau,\bold N\right>+
\frac{B}{|\bold v|}\cdot\partial_t\psi+\frac{B^2}{|\bold v|^2}
\varphi\endsplit\tag3.15
\\
\split
\partial_t\left<\partial_t\boldsymbol\tau,\bold M\right>&=
\left<\partial_{tt}\boldsymbol\tau,\bold M\right>+
\left<\partial_t\boldsymbol\tau,\partial_t\bold M\right>=\\
&=\left<\partial_{tt}\boldsymbol\tau,\bold M\right>-
\frac B{|\bold v|}\partial_t\varphi+\frac{B^2}{|\bold v|^2}
\psi\endsplit\tag3.16
\endalign
$$
Differentiating the right hand sides in the same equations
\thetag{3.14} and taking into account \thetag{3.10},
\thetag{3.11} and \thetag{3.12} we have
$$
\gather
\split
\partial_t\left<\partial_t\boldsymbol\tau,\bold N\right>&=
\partial_{tt}\varphi-\frac{B}{|\bold v|}\partial_t\psi+\frac
{BA}{|\bold v|^2}\psi-\\
&-\beta_1 \psi-\beta_3\frac{A}{|\bold v|}\psi-
\beta_4\frac{B}{|\bold v|}\psi
\endsplit\tag3.17\\
\split
\partial_t\left<\partial_t\boldsymbol\tau,\bold M\right>&=
\partial_{tt}\psi+\frac{B}{|\bold v|}\partial_t\varphi-
\frac{BA}{|\bold v|^2}\varphi+\\
&+\beta_1\varphi+\beta_3\frac{A}{|\bold v|}\varphi+
\beta_4\frac{B}{|\bold v|}\varphi
\endsplit\tag3.18
\endgather
$$
The second derivative of the vector $\boldsymbol\tau$ in
\thetag{3.15} and \thetag{3.16} can be calculated on the base of
\thetag{3.7}, \thetag{3.8} and \thetag{3.9}. Then we get
$$
\gather
\split
\left<\partial_{tt}\boldsymbol\tau,\bold N\right>&=
\alpha_1\varphi+\alpha_2\psi+
\alpha_3\left(\partial_t\varphi-\frac{B}{|\bold v|}\varphi\right)+
\\
&+\left(\alpha_4-\frac{B}{|\bold v|}\right)
\cdot\left(\partial_t\psi+\frac{B}{|\bold v|}\varphi\right)
\endsplit\tag3.19
\\
\split
\left<\partial_{tt}\boldsymbol\tau,\bold M\right>&=
\beta_1\varphi+\beta_2\psi+
\beta_3\left(\partial_t\varphi-\frac{B}{|\bold v|}\psi\right)+
\\
&+\left(\beta_4+\frac{A}{|\bold v|}\right)
\cdot\left(\partial_t\psi+\frac B{|\bold v|}\varphi\right)
\endsplit\tag3.20
\endgather
$$
As a result of equating \thetag{3.15} with \thetag{3.17} and
\thetag{3.16} with \thetag{3.18} after substituting \thetag{3.19}
and \thetag{3.20} for $\varphi$ and $\psi$ we obtain
$$
\align
&\aligned
&\partial_{tt}\varphi-\frac{B}{|\bold v|}\partial_t\psi
+\frac{BA}{|\bold v|^2}\psi-\beta_1\psi-
\beta_3\frac{A\psi}{|\bold v|}-\beta_4\frac{B\psi}{|\bold v|}=\\
&=\alpha_1\varphi+\alpha_2\psi+\alpha_3\left(\partial_t\varphi-
\frac{B}{|\bold v|}\psi\right)+\alpha_4\left(\partial_t\psi+
\frac{B}{|\bold v|}\varphi\right)
\endaligned\tag3.21\\
&\aligned
&\partial_{tt}\psi+\frac{B}{|\bold v|}\partial_t\varphi-
\frac{BA}{|\bold v|^2}\psi+\beta_1\varphi+
\beta_3\frac{A\varphi}{|\bold v|}+
\beta_4\frac{B\varphi}{|\bold v|}=\\
&=\beta_1\varphi+\beta_2\psi+
\left(\beta_3-\frac{B}{|\bold v|}\right)\cdot
\left(\partial_t\varphi-\frac{B}{|\bold v|}\psi\right)+\\
&+\alpha_4\left(\beta_4+\frac{A}{|\bold v|}\right)\cdot
\left(\partial_t\psi+\frac{B}{|\bold v|}\varphi\right)
\endaligned\tag3.22
\endalign
$$
\par
     Let the dynamical system \thetag{1.1} with the force field
\thetag{3.2} be accepting the normal shift in $\Bbb R^2$.
Submanifolds of codimension 1 in $\Bbb R^2$ are the plane curves.
It is convenient to use the natural parameter on them measuring
the arc length referenced to some fixed point on the curve. Let
the parameter $s$ in the Cauchy problem initial data \thetag{3.6}
do coincide with the natural parameter on the curve. Then for the
transformations of the normal shift these initial data may be
rewritten as follows
$$
\left.\bold r(t,s)\right|_{t=0}=\bold r(s)\hskip 10em\left.
\partial_t\bold r(t,s)\right|_{t=0}=v(s)\bold n(s)\tag3.23
$$
The derivative $\partial_s\bold r(s)=\boldsymbol\tau(s)$ is the
unit tangent vector for the curve while $\bold n(s)$ is the unit
normal vector for it. \par
     From \thetag{3.23}, \thetag{3.14} and from the orthogonality
of the vectors $\boldsymbol\tau(s)$ and $\bold n(s)$ we derive
the following initial data for the function $\varphi$ introduced
in $\thetag{3.13}$
$$
\left.\varphi\right|_{t=0}=0\hskip 10em\left.\partial_t\varphi
\right|_{t=0}=\partial_sv(s)+\frac B{v(s)}\tag3.24
$$
By the proper choice of the function $v(s)$ in \thetag{3.23} and
\thetag{3.24} (see also the definition 2) one can vanish the
right hand side of the second initial data statement in
\thetag{3.24}. Such proper choice is defined by the following
differential equation
$$
v'_s=-\frac{B(\bold r(s),v(s)\bold n(s))}{v(s)}\tag3.25
$$
The right hand side of \thetag{3.25} as well as the function
$v(s)$ itself then depend on the form of the curve. Once the
choice \thetag{3.25} for $v(s)$ is made the initial data
\thetag{3.14} take the form
$$
\left.\varphi\right|_{t=0}=0\hskip 10em\left.\partial_t\varphi
\right|_{t=0}=0\tag3.26
$$
Now let's come back to the differential equations \thetag{3.21}
and \thetag{3.22}. They are linear with respect to $\varphi$ and
$\psi$ and altogether they form the complete system of the
differential equations. As for the initial data \thetag{3.26}
they are not enough to determine the Cauchy problem for this
system. However one can make the special choice of the functions
$A$ and $B$ in \thetag{3.2} such that the coefficients of
$\psi$ and $\partial_t\psi$ in \thetag{3.21} vanish making
\thetag{3.21} the separate equation of the second order with
respect to the function $\varphi$. Because of the linearity the
Cauchy problem \thetag{3.26} for this equation has the unique
solution being identically zero. In this case vectors
$\boldsymbol\tau$ and $\bold N$ are perpendicular for all $s$ and
$t$ which is just the condition of normal shift. The proper
choice of $A$ and $B$ leading to this case is described by the
following theorem.
\proclaim{Theorem 1} Dynamical system \thetag{1.1} with the right
hand side of the form \thetag{3.2} is the system accepting the
normal shift in $\Bbb R^2$ if and only if the following
relationships
$$
\align
&B=-|\bold v|\alpha_4\tag3.27
\\
&\frac{BA}{|\bold v|^2}-\beta_1-\beta_3\frac A{|\bold v|}-
\beta_4\frac B{|\bold v|}=\alpha_2-\alpha_3\frac B{|\bold v|}
\tag3.28
\endalign
$$
are fulfilled. Here $\alpha_1$, $\alpha_2$, $\alpha_3$,
$\alpha_4$ and $\beta_1$, $\beta_2$, $\beta_3$, $\beta_4$ are the
coefficients in \thetag{3.4} and \thetag{3.5}.
\endproclaim
\par
\head
4. Some examples in $\Bbb R^2$.
\endhead
     Because of the derivatives in \thetag{3.4} and \thetag{3.5}
the equations \thetag{3.27} and \thetag{3.28} are the system of
the partial differential equations with respect to $A$ and $B$
defining in turn the force field $\bold F$ according to
\thetag{3.2}. In order to make these equations more explicit one
should use the explicit form of the vectors $\bold N$ and
$\bold M$
$$
\bold N=\frac{1}{|\bold v|}\Vmatrix v^1 \\ v^2 \endVmatrix
\hskip 10em
\bold M=\frac{1}{|\bold v|}
\Vmatrix\format \r\\ -v^2 \\ v^1 \endVmatrix
$$
For the function $B$ then from \thetag{3.27} we may obtain its
expression via the function $A$
$$
B=v^2\frac{\partial A}{\partial v^1}-
v^1\frac{\partial A}{\partial v^2}\tag4.1
$$
Further substitution of \thetag{4.1} into \thetag{3.5} and
\thetag{3.28} let us bring the system of equation \thetag{3.27}
and \thetag{3.28} to the form of one equation with respect to
$A$. We do not write this equation here because it is very huge.
But in place of it we consider some examples when this equation
can be simplified to rather observable size. \par
     {\bf Case 1.} Spacially homogeneous force field directed
along the vector $\bold v$ of velocity. Function $A$ in this
case does not depend on $\bold r$ and $B$ is equal to zero
identically. From \thetag{4.1} then we obtain that $A$ depends
only on the modulus of velocity $A=A(|\bold v|)$. Therefore
$\alpha_2=0$ and the equation \thetag{3.28} become the identity.
Note that this case is trivial from geometrical point of view
since trajectories  of particles are straight lines and
associated transformation in \thetag{2.1} coincides with the
classical normal shift. \par
     {\bf Case 2.} Spacially homogeneous  but anisotropic force
field. Both functions $A$ and $B$ do not depend on $\bold r$.
Let's denote $v=|\bold v|$ the modulus of velocity and denote via
$\theta$ the angle between $\bold v$ and some fixed direction in
space. Then $B=-\partial_\theta A=-A_\theta$ and the equation
\thetag{3.28} is written as follows
$$
AA_{\theta}-vAA_{\theta v}+A_{\theta}A_{\theta\theta}=
-vA_{\theta}A_{v}\tag4.2
$$
Equation \thetag{4.2} has the particular solution with the
separated variables $A=A(v)\cos(\theta)$. Force field then is
of the form
$$
\bold F = A(v)\bold N\cos(\theta) + A(v)\bold M \sin(\theta)
\tag4.3
$$
The modulus of force here $|\bold F|=A(v)$ is some arbitrary
function of $v$. The direction of force $\bold F$ form with the
fixed direction in space the angle $2\theta$ twice as greater
than the angle between $\bold v$ and that direction
(see fig. 1). \par
\vskip 1em
\centerline{\boxed{\quad Place\quad for\quad fig. 1.\quad}\qquad
\boxed{\quad Place\quad for\quad fig. 2.\quad}}
\vskip 1em
\par
     {\bf Case 3.} Spacially unhomogeneous force field with the
central point. Here it is convenient to introduce new variables
$\rho$, $\gamma$, $v$ and $\theta$ according to the fig. 2
$$
\aligned
r^1&=\rho\cos(\gamma) \\
r^2&=\rho\sin(\gamma)
\endaligned
\hskip 5em
\aligned
v^1&=v\cos(\gamma+\theta) \\
v^2&=v\sin(\gamma+\theta)
\endaligned
\tag4.4
$$
{}From \thetag{4.4} it is not difficult to derive the following
formulae for different functions in the equations \thetag{3.27}
and \thetag{3.28}
$$
\alignat 3
\alpha_2&=-\sin(\theta)A_\rho+\dfrac{\cos(\theta)}{\rho}(A_
\gamma-A_\theta) &\qquad \alpha_4&=\frac1vA_\theta &\qquad
\alpha_3&=A_v\\ \beta_1&=\cos(\theta)B_\rho+
\dfrac{\sin(\theta)}{\rho}(B_\gamma-B_\theta) &\qquad \beta_4&=
\frac1vB_\theta &\qquad \beta_3&=B_v
\endalignat
$$
Let's substitute these expressions into \thetag{3.27} and
\thetag{3.28} and impose one additional condition  $A_\gamma=0$
(the condition of isotropy for all rays coming out from the
central point). As a result we obtain the following equations
$$
\aligned
&B=-A_\theta
\\
&-\frac{AA_\theta}{v^2}+\cos(\theta)A_{\theta\rho}-
\frac{\sin(\theta)}{\rho}A_{\theta\theta}+\frac{AA_{\theta v}}v-
\frac{A_\theta A_{\theta\theta}}{v^2}=\\
&=-\sin(\theta)A_\rho-
\frac{\cos(\theta)}{\rho}A_\theta+\frac{A_\theta A_v}v
\endaligned\tag4.5
$$
The solution of the equations \thetag{4.5} with separated
variables here has the following form
$$
A=\frac{A(v)}{\rho}\cos(\theta)\hskip 10em B=\frac{A(v)}{\rho}
\sin(\theta)
$$
where $A(v)$ is some arbitrary function of modulus of velocity
$\bold v$. Corresponding dynamical system has the following force
field
$$
\bold F=\frac{A(v)\bold N\cos(\theta)+A(v)\bold M\sin(\theta)}
{\rho}\tag4.6
$$
(see fig. 2). Note, that both dynamical systems with force fields
\thetag{4.3} and \thetag{4.6} are integrable via quadratures.
Moreover for some special choice of $A(v)$  they are explicitly
integrable. For instance when $A(v)=\const$ the trajectories of
the system \thetag{4.3} are the cycloids.


\Refs
\ref \no 1 \by Bianchi L.
\paper Ricerche sulle superficie a curvatura constante e
   sulle elicoid. \jour Ann. Scuola Norm. Pisa \yr 1879
\vol 2 \page 285
\endref
\ref \no 2 \by Lie S.
\paper Zur Theorie der Fl\"achen konstanter Kr\"ummung, III, IV.
\jour Arch. Math. og Naturvidenskab \yr 1880 \vol  5 \issue 3
\pages 282--306, 328--358 and 398--446
\endref
\ref \no 3 \by B\"acklund A.V.
\paper Om ytor med konstant negativ kr\"okning.
\jour  Lunds Universitets \AA rs-skrift, \yr 1883 \vol 19
\endref
\ref \no 4\by Darboux G.
\book Le\c cons sur la th\'eorie g\'en\'erale des surfaces, III.
\publ Gauthier-Villars et Fils \publaddr Paris \yr 1894
\endref
\ref \no 5\by Ibragimov N.Kh.\book Groups of transformations in
Mathematical Physics.
\publ Nauka \publaddr Moscow \yr 1983
\endref
\ref \no 6 \by Olver P.J.
\book Application of Lie Groups to Differential Equation.
\publ Springer-Verlag
\endref
\ref \no 7 \by Teneblat K. and Terng C.L.
\paper B\"acklund theorem  for $n$-dimensional submanifolds
of \ $\Bbb R^{2n-1}$.
\jour Annals of Math. \yr 1980 \vol 111 \issue 3
\pages 477-490
\endref
\ref \no 8 \by Terng C.L.
\paper  A higher dimensional generalization of Sine-Gordon
   equation and its soliton theory. \jour Annals of Math. \yr 1980
   \vol 111 \issue 3 \pages 491--510
\endref
\ref \no 9 \by Chern S.S. and Terng C.L.
\paper An analogue of B\"acklund theorem in affine geometry.
\jour Rocky Mount Journ. of Math. \yr 1980 \vol 10 \issue 1
\page 105
\endref
\ref \no 10 \by Bianchi L.
\paper Sopra le deformazioni isogonali delle superficie a
    curvatura constante in geometria elliptica ed hiperbolica.
\jour Annali di Matem. \yr 1911 \vol 18 \issue 3 \pages 185--243
\endref
\ref \no 11 \by Teneblat K.
\paper B\"acklund theorem for submanifolds of space forms
    and a generalized wave equation.  \jour  Bol. Soc. Bras. Math.
    \yr 1985 \vol 18 \issue 2 \pages 67--92
\endref
\endRefs
\enddocument